# Spectroscopy of nickel monosulfide in 450 − 560 nm by laser-induced fluorescence and dispersed fluorescence techniques

Li Wang, Junfeng Zhen, Jianqiang Gao, Qun Zhang*, Yang Chen*

*Hefei National Laboratory for Physical Sciences at the Microscale and Department of Chemical Physics, University of Science and Technology of China, Hefei, Anhui 230026, People's Republic of China*

*Authors to whom correspondence should be addressed. Electronic addresses: qunzh@ustc.edu.cn (Q. Zhang) and yangchen@ustc.edu.cn (Y. Chen)

A B S T R A C T

Laser-induced fluorescence excitation spectrum of NiS in the wavelength range 450 − 560 nm has been recorded and analyzed. Thirty-five vibronic transition bands have been observed, twenty-nine of which are reported for the first time. Rotational analyses indicated that the observed bands can be attributable to the $[\Omega = 0] - X^3\Sigma_0^-$ transitions of $^{58}$NiS (and $^{60}$NiS). Twenty-five bands have been suggested to be grouped into four vibrational progressions. Furthermore, through dispersed fluorescence measurements we directly obtained the energies for the ground-state vibrational levels up to $\upsilon'' = 6$ as well as the vibrational frequency and the anharmonicity constant for the ground state of $^{58}$NiS.

# 1. Introduction

Transition metal oxides and sulfides have attracted much attention due to their fundamental importance in many fields, such as catalysis [1], materials science [2], high-temperature chemistry [3], and astrophysics [4]. Transition-metal oxides are commonly used as versatile catalyst in industry; however, non-specific product formation occurs in some processes due to their high reactivity. In contrast, transition-metal sulfides as catalyst are less reactive, hence susceptible to being poisoned and can show higher selectivity. Moreover, transition-metal sulfides are believed to play a crucial role in biochemistry, for they are found in the reaction centers of many enzymes and have even been considered to be essential for the evolution of life [5].

Over the past years, most of the 3d transition-metal monoxides, such as VO, CrO, MnO, FeO, CoO, NiO, and CuO have been well studied both theoretically and experimentally [6–12]. However, the spectroscopic data for their monosulfide counterparts are still rather incomplete. Such incompleteness is even apparent for nickel monosulfide (NiS), in that only the ground state ($X^3\Sigma^-$) [13] and several excited states (including $^3\Sigma^-$ and $^3\Pi$ states) [14-16] of this species have hitherto been experimentally characterized. By means of source-modulation microwave spectroscopy, Yamamoto *et al.* [13] recently observed the rotational spectrum of NiS in the region between 135 and 314 GHz. Their spectral analyses yielded the molecular constants with high precision, such as the rotational, centrifugal distortion, and several fine-structure constants, for the $X^3\Sigma^-$ ground state of NiS. More recently Ram *et al.* [15] investigated the emission spectrum of NiS in the 2000 – 7500 cm$^{-1}$ region using a Fourier transform spectrometer. They assigned the observed bands to a

$^3\Pi_i - X^3\Sigma^-$ transition analogous to the transition of NiO observed in the near infrared, and the obtained spectroscopic constants for the ground state were found to agree well with the values reported in the microwave study by Yamamoto *et al.* [13]. Our group, prior to these works [13,15], has reported on a spectroscopic study of NiS by means of laser-induced fluorescence (LIF) technique [14], in which we tentatively assigned six vibronic bands newly observed in the wavelength range 495 – 555 nm as the $[17.4]^3\Sigma_0^-(\upsilon'=2-7) - X^3\Sigma_0^-(\upsilon''=0)$ transitions. Nevertheless, we recently realized that such an assignment may be questionable, because (1) the $\omega_e''$ value (322 cm$^{-1}$) used in this assignment [14] was taken from a theoretical prediction [17] which obviously deviates too much from that given in Ref. 13 (507.227(35) cm$^{-1}$), a rather convincing value derived from the microwave measurements; (2) the incorrect usage of the $\omega_e''$ value can therefore result in a misleading $\upsilon'$ numbering of the upper $^3\Sigma_0^-$ electronic state. Furthermore, we detected that the spectroscopic constants for the upper state given in Ref. 14 are also misleading, since the $B_0''$ value (0.182 cm$^{-1}$) [14] used throughout the entire rotational analyses is too small compared to that given in Ref. 13 (6351.4222 MHz, *i.e.*, 0.21186 cm$^{-1}$).

We therefore revisited the LIF excitation spectrum of NiS between 495 and 555 nm as reported in Ref. 14, and further extended the laser excitation wavelengths to a broader range, *i.e.*, 450 – 560 nm in this work. Apart from the six vibrational bands reported in Ref. 14, additional 29 bands of NiS were observed for the first time. Based on a global rotational, isotopic shift, and band interval analysis, we assigned 25 out of the 35 bands in total to four progressions with each corresponding to a $[\Omega=0](\upsilon') - X^3\Sigma_0^-(\upsilon''=0)$ transition, namely, four upper electronic states of NiS were identified in the energy range of ~18100 – 22100

cm$^{-1}$ (*i.e.*, 450 – 560 nm).  As the $B_0''$ value given in Ref. 13 was used throughout the rotational analyses in the present work, reliable rotational constants were obtained for the upper vibronic states related to all the 35 bands, among which those for the six bands reported in Ref. 14 were corrected.  Furthermore, we recorded the dispersed fluorescence (DF) spectra at two differing excitation wavelengths, from which the energies for the ground-state vibrational levels up to $v''= 6$ were determined.  The DF measurements also allowed us to obtain experimentally, for the first time, the vibrational frequency ($\omega_e''$ = 506 ± 5 cm$^{-1}$) and the anharmonicity constant ($\omega_e\chi_e''$ = 2.29 ± 0.65 cm$^{-1}$) for the ground state of $^{58}$NiS.

## 2. Experimental

The experimental apparatus has been described in detail elsewhere [18–20].  Briefly, the NiS molecules were produced by the reaction of H$_2$S molecules with the nickel atoms sputtered from a pair of pure nickel pin electrodes under a pulsed DC discharge condition.  The H$_2$S sample seeded in argon (~2%) at a stagnation pressure of ~5 atm passed through a pulsed nozzle (General Valve Co.) with an orifice diameter of 0.5 mm into the vacuum chamber.  The nickel pins used for DC discharging the H$_2$S/Ar gas was fixed in a Teflon disk, and set at a spacing of ~1 mm.  The background pressure of the vacuum chamber was ~4×10$^{-4}$ and ~3×10$^{-5}$ Torr, with and without operation of the free jet, respectively.

The light source used was a tunable dye laser (Lumonics, HT-500) pumped by a Nd:YAG laser (Spectra Physics, GCR-190).  The output of the pulsed dye laser (linewidth ~0.1 cm$^{-1}$, pulse duration ~5 ns) was introduced into the vacuum chamber and crossed the jet flow perpendicularly about 3 cm downstream from the point of discharge.

The LIF excitation spectra were recorded by monitoring the total fluorescence as a function of laser wavelength. No attempt was made to normalize the spectral intensity against the laser power. The DF spectra were obtained by fixing the probe laser frequency at or near a strong *R*-head and scanning the monochromator (Zolix, Omni-λ300) with a 1-mm-wide slit. The relative time delays among the nozzle, the laser, and the discharge were controlled by a multichannel pulsed delay generator. Laser wavelengths were calibrated by a wavemeter (Coherent, WaveMaster 33-2650).

## 3. Results and discussion

Figure 1 shows a survey LIF excitation spectrum of NiS in the laser wavelength range of 450 – 560 nm. This spectrum is very complex (comprising 35 rotationally analyzable bands), which may result from the presence of huge numbers of the low-lying electronic states and the widespread perturbations among the electronic states; spin multiplicity may add more complexity to the spectrum [14,16]. Apart from the spectral features arising from $^{58}$NiS, those arising from its isotopic molecule $^{60}$NiS were also observed (natural abundance ratio of $^{58}$Ni/$^{60}$Ni ~ 2.6:1). Since all the NiS molecules in the detection zone are in their ground state after having expanded for ∼60 μs from the discharge zone in our experiment, we can conclude that all the vibronic bands observed in this work should arise from the NiS ground state ($X^3\Sigma_0^-$ and $X^3\Sigma_1^-$, see discussion below).

A focus on band-head positions suggests four groupings of bands: (i) a progression from 550.96 nm, namely bands at 550.96, 540.24, 531.01, 521.53, 512.19, 503.67, 494.65, and

485.80 nm.  These bands occur at intervals of ~348 cm$^{-1}$.  The bands labeled (i)'s in the lower panel of Fig. 1 are those observed previously (with similar intervals of ~340 cm$^{-1}$ [14]); (ii) a progression from 508.89 nm, namely bands at 508.89, 499.53, 490.73, 481.90, 473.75, and 465.81 nm.  These bands occur at intervals of ~363 cm$^{-1}$; (iii) a progression from 493.17 nm, namely bands at 493.17, 484.61, 476.62, 468.73, 461.18, and 453.73 nm.  These bands occur at intervals of ~352 cm$^{-1}$; (iv) a progression from 496.21 nm, namely bands at 496.21, 487.54, 479.31, 471.58, and 463.60 nm.  These bands occur at intervals of ~354 cm$^{-1}$.

The 25 vibronic bands grouped into the four progressions are labeled correspondingly (i), (ii), (iii), and (iv) in Fig. 1.  Since perturbations among excited states are widespread [14,16], which leads to irregularities in the isotope shifts as well as the rotational constants obtained from our rotational analyses for these bands, we here adopt the above tentative classification, as did in a recent report on NiO by Balfour *et al.* [11].  We thus hold that the $\upsilon'$ numbering given in Ref. 14 (*i.e.*, the assignments of $[17.4]^3\Sigma_0^-(\upsilon'=2-7)-X^3\Sigma_0^-(\upsilon''=0)$ for the six bands labeled (i)'s in the lower panel of Fig. 1) is doubtful, as has been stated in the Introduction section.

Listed in Table 1 are the observed band heads ($\lambda_{head}$) together with the band origins ($\nu_0$) obtained from the rotational analyses for the 25 bands belonging to the four vibrational progressions.  As the four progressions have been identified (through rotational analyses given below) as being due to the $[\Omega=0](\upsilon')-X^3\Sigma_0^-(\upsilon''=0)$ transition, the vibrational frequencies ($\omega_e'$) of the four newly observed excited states can be estimated to be ~350 cm$^{-1}$; however, the accurate $\omega_e'$ (and $\omega_e\chi_e'$) values for the four excited states cannot be determined in the present work, partly because of the unidentified $v'$ numbering.

It is noteworthy that there remain 10 bands (with appreciable spectral intensity) unlabeled in Fig. 1. Through careful rotational analyses we confirmed that the 10 rotationally analyzable bands, similar to the aforementioned 25 bands (labeled (i) – (iv) in Fig. 1), should also originate from the $[\Omega=0]-X^3\Sigma_0^-$ type transition. Nevertheless, they can neither be classified into the four groupings shown in Fig. 1 nor into any other vibrational progressions; this sort of perturbation-induced spectral complexity can also be found in other 3d transition-metal diatomic molecules (*e.g.*, NiO [11]). Table 2 lists the band heads ($\lambda_{head}$) together with the band origins ($\nu_0$) for the 10 bands not belonging to the four vibrational progressions.

The rotational analyses indicated that all the 35 vibrational bands of NiS show a similar *R-P* rotational structure, as can be seen in the 503.67 nm (~19854.27 cm$^{-1}$) band (*i.e.*, the band labeled (*a*) in Fig. 1) and the 459.54 nm (~21760.89 cm$^{-1}$) band (*i.e.*, the band labeled (*b*) in Fig. 1) shown in Figs. 2 and 3, respectively. The two bands are representative of those in and out of the four vibrational progressions described above, respectively. From Figs. 2 and 3 we can readily detect that both spectra exhibit *P* and *R* (without *Q*) branches and that *P* branch merges with a sharp *R* head. Since the ground state of NiS has been determined to be of $^3\Sigma^-$ symmetry [13], we can reasonably attribute all the observed 35 bands shown in Fig. 1 to the $[\Omega=0]-X^3\Sigma_0^-$ transitions. Very weak though compared to those from $^{58}$NiS, the spectral features arising from $^{60}$NiS are discernible, as shown in Figs. 2 and 3, which in turn confirms that the spectral carrier is indeed NiS.

We note that the rotational temperature is of the order of 50 K, implying that both the $\Omega$

= 0 and $\Omega$ = 1 components would be populated in the ground state. Since the ground-state microwave spectrum yielded a spin-spin constant $\lambda$ = 1087.5 GHz (~36.3 cm$^{-1}$) [13], the spin-spin splitting due to such a quite large $\lambda$ value would be observable within our spectral resolution. This is indeed the case, as we really observed other three sub-bands (*i.e.*, the $[\Omega=0]-X^3\Sigma_1^-$, $[\Omega=1]-X^3\Sigma_1^-$, and $[\Omega=1]-X^3\Sigma_0^-$ bands) than the $[\Omega=0]-X^3\Sigma_0^-$ sub-band (see the schematic transition diagram depicted in Fig. 4). As an example, we show also in Fig. 4 the spectrum that corresponds to that shown in Fig. 3 but has been extended to the red and blue regions. As clearly seen in Fig. 4, the separations between (a) and (c) and between (b) and (d) are both ~71 cm$^{-1}$, in excellent agreement with the ground-state spin-spin splitting value ($2\lambda$ ~ 72.6 cm$^{-1}$ [13]), which further confirms that the rotational assignments given in this work are indeed due to the $[\Omega=0]-X^3\Sigma_0^-$ transitions. Notably, however, owing to the severe spectral blending and the weak LIF intensities of most of these sub-bands in our spectra, it is difficult for us at this stage to give a complete and meaningful analysis for every sub-band adjacent to the most intense $[\Omega=0]-X^3\Sigma_0^-$ sub-band. We here list in Table 3 the band heads (in cm$^{-1}$) for the observed 35 bands (the last 10 rows are those not belonging to the (i) – (iv) progressions), with (a), (b), (c), and (d) corresponding respectively to the $[\Omega=0]-X^3\Sigma_1^-$, $[\Omega=1]-X^3\Sigma_1^-$, $[\Omega=0]-X^3\Sigma_0^-$, and $[\Omega=1]-X^3\Sigma_0^-$ sub-bands (refer to the schematic transition diagram shown in Fig. 4). The missing sub-bands in Table 3 are due to the severe spectral blending or weak signals in this spectral region. From Table 3 we can see that the values of [(c) - (a)] and [(d) - (b)] (both correlating to the ground-state spin-spin splitting) are around 70 cm$^{-1}$, while those of [(b) - (a)] and [(d) - (c)] (both correlating to the excited-state spin-spin splitting) range from 3 to 40

cm$^{-1}$. This observation may result from the fact that much stronger perturbations occur in the excited states than in the ground state, which in turn leads to the spectral complexity showing irregular separations of the two upper-state $\Omega$ components for different vibrational bands. Admittedly, the upper-state splitting values given in Table 3 are more suggestive than indicative.

The upper part of Figs. 2 and 3 shows the observed spectrum of both $^{58}$NiS and $^{60}$NiS with *P* and *R* branches indicated by ticks showing corresponding *J* numbers. Table 4 lists rotational assignments for the 503.67 nm and 459.54 nm bands, the bands labeled (*a*) and (*b*) in Fig. 1, respectively. Note that the rotational lines with *J* > 25 in Figs. 2 and 3 can rarely be measured because of the jet-cooled conditions in our experiment. A least-squares fit using the PGOPHER program [21] allowed us to simulate the rotational spectra. The simulated spectra (shown in the lower part of Figs. 2 – 4) match nicely with the observed ones, which further supports our spectral assignments. It is worth noting that both spectra presented in Figs. 2 and 3 show bands where the *P* and *R* branches overlap almost perfectly, so each line in the spectrum is made up of a *P* and an *R* line (*e.g.*, *P*(1) and *R*(8) are coincident for the 459.54 nm band). In such cases, particularly with the presence of the band head, it is very difficult to identify the first line of a branch within our spectral resolution. Fortunately, with the aid of the rotational simulations we can reasonably argue that the rotational line of the *P*(*J*) branch starts from *J* = 1 for both spectra shown in Figs. 2 and 3, because otherwise we can hardly make the simulated and observed spectra match nicely as shown in Figs. 2 and 3. Our rotational assignments, on the other hand, imply that these bands arise from the $\Omega = 0 - \Omega = 0$ type transition; we have to admit, however, that such an implication based on the correctness of $J$ numbering cannot be used unambiguously as a means to pin-point the $\Omega$ value discussed here. Nevertheless, the different pattern of intensities for the observed bands (as clearly shown in Fig. 4) are sufficiently distinctive to assign the four sub-bands, *i.e.*, $[\Omega=0]-X^3\Sigma_1^-$, $[\Omega=1]-X^3\Sigma_1^-$, $[\Omega=0]-X^3\Sigma_0^-$, and $[\Omega=1]-X^3\Sigma_0^-$, as discussed above.

The data derived from the rotational analyses for the $[\Omega=0]-X^3\Sigma_0^-$ sub-band, including the band origins ($\nu_0$), the rotational constants ($B'$), and the isotope shifts ($\Delta\nu_0^i = \nu_0(^{58}\text{NiS}) - \nu_0(^{60}\text{NiS})$), are listed in Tables 1 and 2. Note that (1) for each of the four vibrational progressions, the $\Delta\nu_0^i$ values given in Table 1 do not exhibit good linear relationship with band positions; and (2) there remain some vacancies for the $\Delta\nu_0^i$ values in Table 2. These may be interpreted in terms of widespread perturbations among excited states [11,14,16] and perturbation-induced spectral congestion.

In addition to the above LIF work pertinent to the newly observed excited states of NiS in the wavelength range of 450 – 560 nm, we further recorded the dispersed fluorescence (DF) spectra at two differing excitation wavelengths in an attempt to directly acquire information about the $X^3\Sigma^-$ ground state of NiS. In two of the progressions identified above, the lead bands, specifically those at 503.67 nm (labeled (*a*) in Fig. 1) and 476.62 nm (labeled (*c*) in Fig. 1), are among the strongest members. DF spectra were recorded with probe laser excitation fixed to coincide with the most intense $R$-head in the $^{58}$NiS isotopomer of each band.

Figure 5, as an example, shows the obtained DF spectrum with excitation wavelength ($\lambda_{exc}$) fixed at 476.62 nm. The displacements of the seven peaks in Fig. 5, with respect to the position marked with an arrow, read 0, 505, 1002, 1495, 1988, 2479, and 2952 cm$^{-1}$, respectively. [Note: The DF spectrum recorded with $\lambda_{exc}$ = 503.67 nm was qualitatively similar.] The predominant emission was found at ~476.62 nm, exactly equivalent to the wavelength used for excitation, which implies that this strongest peak should correspond to the $[\Omega=0](\upsilon') - X^3\Sigma_0^-(\upsilon''=0)$ transition. We thereby anticipate that the remaining six peaks may correspond to the $[\Omega=0](\upsilon') - X^3\Sigma_0^-(\upsilon''=1-6)$ transitions. Fitting the above displacement values into the customary equation $G(\upsilon'') = \omega_e''(\upsilon''+\frac{1}{2}) - \omega_e x_e''(\upsilon''+\frac{1}{2})^2$ yields the vibrational frequency ($\omega_e''$ = 506 ± 5 cm$^{-1}$) as well as the anharmonicity constant ($\omega_e \chi_e''$ = 2.29 ± 0.65 cm$^{-1}$) for the ground state of $^{58}$NiS. These two vibrational constants we obtained directly from our DF measurements turned out to agree reasonably with the calculated ones ($\omega_e''$ = 507.227 ± 0.035 cm$^{-1}$ and $\omega_e \chi_e''$ = 2.20341 ± 0.00038 cm$^{-1}$) that were based on the microwave spectroscopic data ($B_e''$, $D_e''$, and $\alpha_e''$) [13] and usage of the equations $\omega_e'' = \sqrt{\frac{4B_e''^3}{D_e''}}$ and $\omega_e x_e'' = B_e''(\frac{\alpha_e'' \omega_e''}{6B_e''^2}+1)^2$. Such an agreement further supports our LIF work described above. To the best of our knowledge, the DF work presented here represents a first direct measurement of the vibrational energies and the two vibrational constants ($\omega_e''$ and $\omega_e \chi_e''$) for the $X^3\Sigma^-$ ground state of $^{58}$NiS.

## 4. Conclusions

The jet-cooled laser-induced fluorescence excitation spectrum of NiS has been investigated between 450 and 560 nm where thirty-five vibronic transition bands have been observed and assigned as the $[\Omega=0] - X^3\Sigma_0^-$ transitions based on our rotational and isotopic shift analyses. Four vibrational progressions involving twenty-five bands are suggested.

Dispersed fluorescence spectra we recorded at two differing excitation wavelengths map the ground-state vibrational levels up to $\upsilon''=6$, from which the experimentally derived valves of the $^{58}$NiS ground-state vibrational frequency ($\omega_e'' = 506 \pm 5$ cm$^{-1}$) and anharmonicity constant ($\omega_e\chi_e'' = 2.29 \pm 0.65$ cm$^{-1}$) have been given for the first time.

## Acknowledgements

This work was supported by the National Natural Science Foundation of China (Grant Nos. 20673107 and 20873133), the Ministry of Science and Technology of China (Grant Nos. 2007CB815203 and 2010CB923302), the Chinese Academy of Sciences (KJCX2-YW-N24), and the Scientific Research Foundation for the Returned Overseas Chinese Scholars, Ministry of Education of China.  The authors enjoyed helpful discussions with the anonymous reviewer.

# References

[1] G.D. Cody, Annu. Rev. Earth Planet. Sci. 32 (2004) 569.

[2] C.W. Bauschlicher Jr., P. Maitre, Theor. Chim. Acta 90 (1995) 189.

[3] H. Spinrad, R.F. Wing, Annu. Rev. Astron. Astrophys. 7 (1969) 249.

[4] A.J. Merer, Annu. Rev. Phys. Chem. 40 (1989) 407.

[5] I. Kretzschmar, D. Schröder, H. Schwarz, P.B. Armentrout, Advances in Metal and Semiconductor Clusters – Metal-Ligand Bonding and Metal-Ion Solvation 5 (2001) 347.

[6] L. Karlsson, B. Lindgren, C. Lundevall, U. Sassenberg, J. Mol. Spectrosc. 181 (1997) 274.

[7] M. Barnes, P.G. Hajigeorgiou, A.J. Merer, J. Mol. Spectrosc. 160 (1993) 289.

[8] K.M. Green, R.P. Kampf, J.M. Parson, J. Chem. Phys. 112 (2000) 1721.

[9] M. Barnes, M.M. Fraser, P.G. Hajigeorgiou, A.J. Merer, S.D. Rosner, J. Mol. Spectrosc. 170 (1995) 449.

[10] M. Barnes, D.J. Clouthier, P.G. Hajigeorgiou, G. Huang, C.T. Kingston, A.J. Merer, G.F. Metha, J.R.D. Peers, S.J. Rixon, J. Mol. Spectrosc. 186 (1997) 374.

[11] W.J. Balfour, J. Cao, R.H. Jensen, R. Li, Chem. Phys. Lett. 385 (2004) 239.

[12] O. Appelblad, A. Lagerqvist, I. Renhorn, R.W. Field, Phys. Scr. 22 (1981) 603.

[13] T. Yamamoto, N. Tanimoto, T. Okabayashi, Phys. Chem. Chem. Phys. 9 (2007) 3744.

[14] X.F. Zheng, T.T. Wang, J.R. Guo, C.X. Chen, Y. Chen, Chem. Phys. Lett. 394 (2004) 137.

[15] R.S. Ram, S. Yu, I. Gordon, P.F.Bernath, J. Mol. Spectrosc. 258 (2009) 20.

[16] J.F. Zhen, L. Wang, C.B. Qin, Q. Zhang, Y. Chen, Chin. J. Chem. Phys. 22 (2009) 668.

[17] A.I. Boldyrev, J. Simons, Periodic Table of Diatomic Molecules (Wiley, New York, 1997).

[18] Y. Chen, J. Jin, C.J. Hu, X.L. Yang, X.X. Ma, C.X. Chen, J. Mol. Spectrosc. 203 (2000) 37.

[19] J. Jin, Q. Ran, X.L. Yang, Y. Chen, C.X. Chen, J. Phys. Chem. A 105 (2001) 11177.

[20] X.P. Zhang, J.R. Guo, T.T. Wang, L.S. Pei, Y. Chen, C.X. Chen, J. Mol. Spectrosc. 220 (2003) 209.

[21] C.M. Western, PGOPHER: A Program for Simulating Rotational Structure (Univ. of Bristol, http://pgopher.chm.bris.ac.uk).

# **Figure Captions**

Fig. 1.   The survey LIF excitation spectrum of NiS in the wavelength range of 450 − 560 nm.   The 25 bands labeled (i), (ii), (iii), and (iv) are those can be grouped into four progressions, while the remaining 10 bands are not labeled.   [See text for details.]   The two bands indicated by (*a*) and (*b*) are presented in more detail in Figs. 2 and 3, respectively.   (*c*) indicates the band whose band head is chosen to be the excitation wavelength (476.62 nm) for the DF measurement shown in Fig. 5.

Fig. 2.   Rotationally resolved LIF excitation spectrum of NiS near 503.67 nm (*i.e.*, the band labeled (*a*) in Fig. 1).   The upper trace is the observed spectrum; while the lower trace is the simulated spectrum.

Fig. 3.   Rotationally resolved LIF excitation spectrum of NiS near 459.54 nm (*i.e.*, the band labeled (*b*) in Fig. 1).   The upper trace is the observed spectrum; while the lower trace is the simulated spectrum.

Fig. 4.   Schematic transition diagram showing the origin of four sub-bands and the rotationally resolved LIF excitation spectrum (observed and simulated) of NiS near 459.54 nm (*i.e.*, the band labeled (*b*) in Fig. 1).   This spectrum corresponds to that shown in Fig. 3 but has been extended to the red and blue regions, showing clearly the other three sub-bands. [See text for details.]

Fig. 5.   The DF spectrum of $^{58}$NiS obtained using a probe laser wavelength fixed at 476.62 nm, which corresponds to the band-head position of the band labeled (*c*) in Fig. 1.   This spectrum can be readily interpreted in terms of emission solely into the vibrational manifold ($v'' = 0 - 6$  as indicated by ticks) of the electronic ground state.

# Table Legends

Table 1

$^{58}$NiS band heads ($\lambda_{head}$, in nm), band origins ($\nu_0$, in cm$^{-1}$), rotational constants ($B'$, in cm$^{-1}$), and $^{58}$NiS/$^{60}$NiS isotopic shifts ($\Delta\nu_0^i$, in cm$^{-1}$) for the 25 bands belonging to the four vibrational progressions (labeled (i) – (iv); see text). Several $\Delta\nu_0^i$ values with question mark remain doubtful (see discussion in the text). The 1σ errors are given in parentheses in units of the last digits quoted.

Table 2

$^{58}$NiS band heads ($\lambda_{head}$, in nm), band origins ($\nu_0$, in cm$^{-1}$), rotational constants ($B'$, in cm$^{-1}$), and $^{58}$NiS/$^{60}$NiS isotopic shifts ($\Delta\nu_0^i$, in cm$^{-1}$) for the 10 bands not belonging to the four vibrational progressions (see text). The 1σ errors are given in parentheses in units of the last digits quoted.

Table 3

$^{58}$NiS band heads (in cm$^{-1}$) for the observed 35 bands (the last 10 rows are those not belonging to the (i) – (iv) progressions, see text). (a), (b), (c), and (d) correspond to the $[\Omega=0]-X^3\Sigma_1^-$, $[\Omega=1]-X^3\Sigma_1^-$, $[\Omega=0]-X^3\Sigma_0^-$, and $[\Omega=1]-X^3\Sigma_0^-$ sub-bands, respectively; refer to the schematic transition diagram shown in Fig. 4.

Table 4

Rotational assignments (in cm$^{-1}$) in the 503.67 and 459.54 nm bands of $^{58}$NiS and $^{60}$NiS.

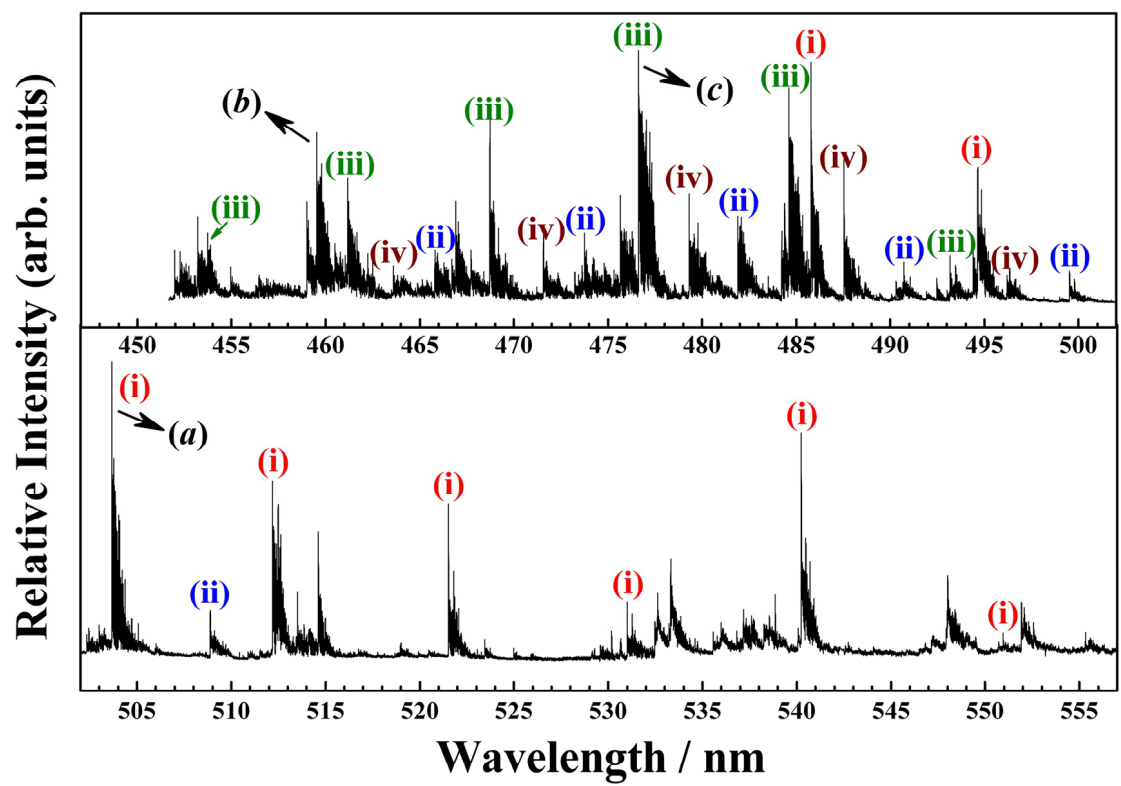

**Fig. 2 (Wang *et al.*)**

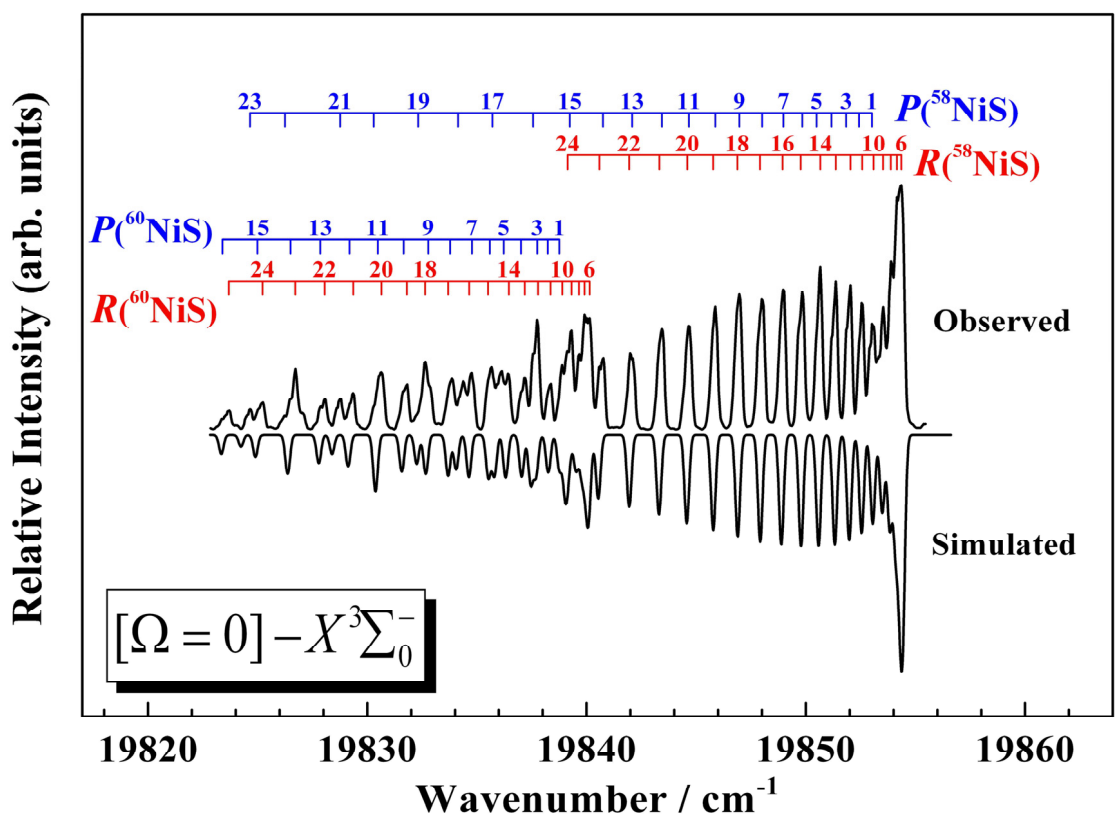

Fig. 3 (Wang *et al.*)

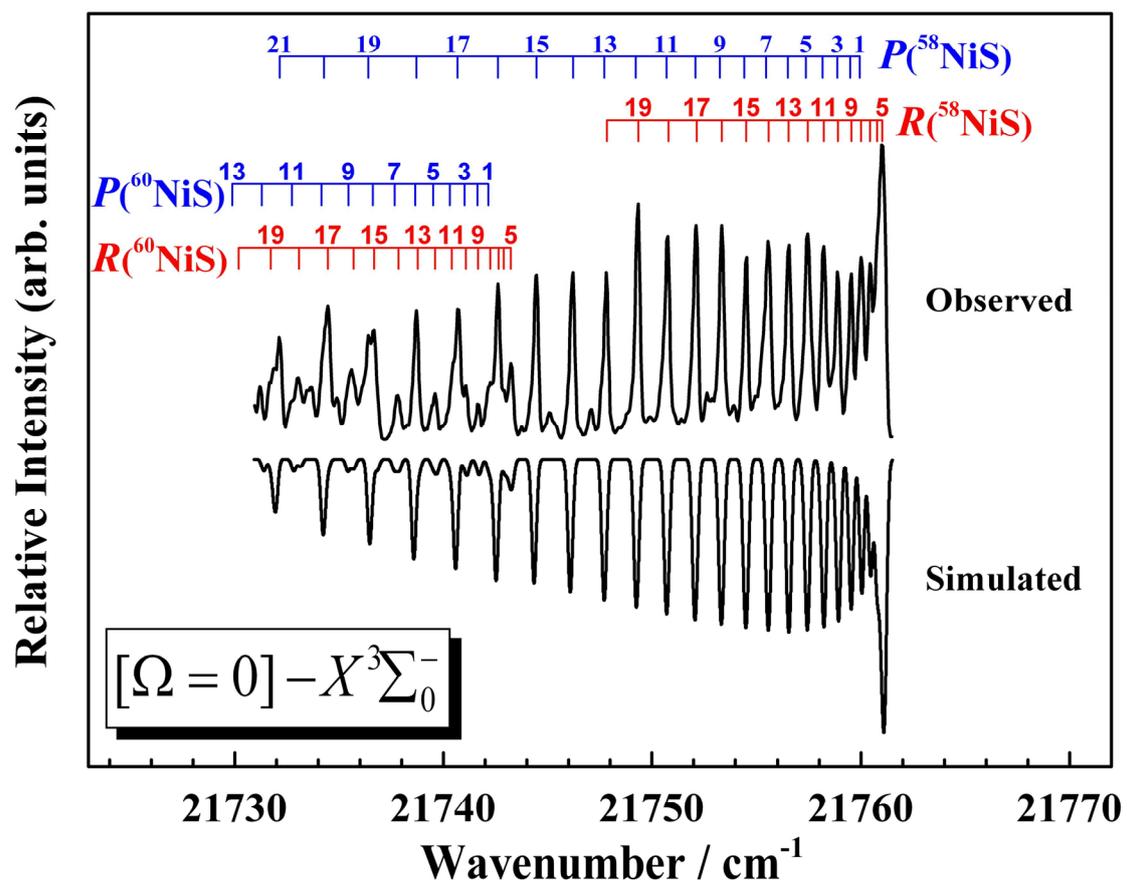

**Fig. 4 (Wang *et al.*)**

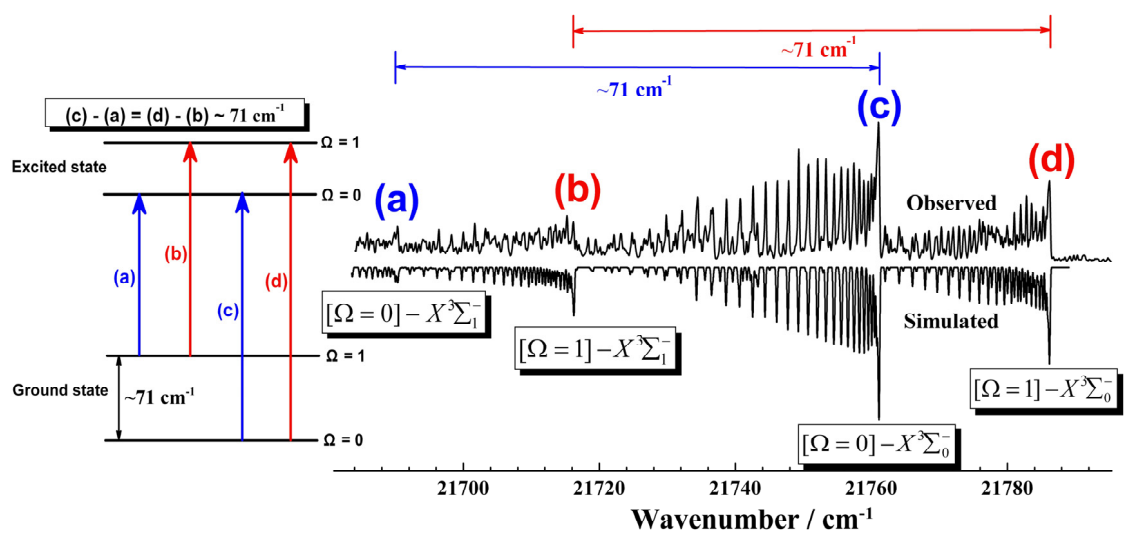

**Fig. 5 (Wang *et al.*)**

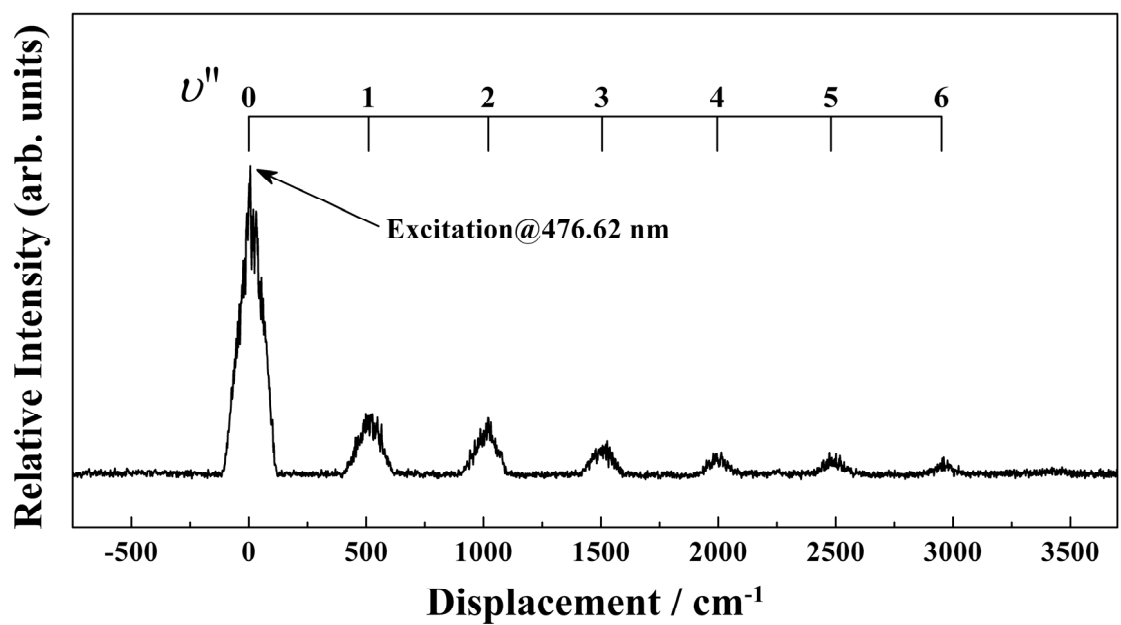

**Table 1**

| $\lambda_{head}$ (nm) | $\nu_0$ (cm$^{-1}$) | $B'$ (cm$^{-1}$) | $\Delta\nu_0^i$ (cm$^{-1}$) | Tentatively assigned progression |
|---|---|---|---|---|
| 550.96 | 18149.50(2) | 0.1735(2) | 4.54(3) | (i) |
| 540.24 | 18509.28(4) | 0.1735(3) | 8.10(5) | (i) |
| 531.01 | 18831.25(3) | 0.1728(2) | 9.56(4) | (i) |
| 521.53 | 19173.56(2) | 0.1737(1) | 10.46(3) | (i) |
| 512.19 | 19523.18(3) | 0.1734(1) | 20.53(4)? | (i) |
| 503.67 | 19853.54(1) | 0.1734(2) | 14.31(2) | (i) |
| 494.65 | 20215.52(3) | 0.1703(2) | 7.70(4)? | (i) |
| 485.80 | 20583.96(3) | 0.1700(2) | 16.71(4) | (i) |
| 508.89 | 19649.60(2) | 0.1711(1) | 8.35(3) | (ii) |
| 499.53 | 20018.03(2) | 0.1679(2) | 10.56(3) | (ii) |
| 490.73 | 20377.40(2) | 0.1665(2) | 14.31(3) | (ii) |
| 481.90 | 20750.58(2) | 0.1650(2) | 13.48(3) | (ii) |
| 473.75 | 21107.51(2) | 0.1696(1) | 19.20(3) | (ii) |
| 465.81 | 21467.34(2) | 0.1625(2) | 11.27(3) | (ii) |
| 493.17 | 20276.27(3) | 0.1682(2) | 12.39(4) | (iii) |
| 484.61 | 20634.16(2) | 0.1700(1) | 20.26(3)? | (iii) |
| 476.62 | 20980.08(2) | 0.1711(1) | 16.60(3) | (iii) |
| 468.73 | 21333.45(3) | 0.1697(3) | 17.07(4) | (iii) |
| 461.18 | 21682.67(2) | 0.1697(1) | 23.07(3) | (iii) |
| 453.73 | 22038.87(3) | 0.1705(2) | 12.81(4)? | (iii) |
| 496.21 | 20151.96(2) | 0.1694(1) | 16.08(3) | (iv) |
| 487.54 | 20510.30(3) | 0.1709(1) | 17.86(4) | (iv) |
| 479.31 | 20862.52(3) | 0.1697(2) | 20.15(4)? | (iv) |
| 471.58 | 21204.61(4) | 0.1688(4) | 16.18(5) | (iv) |
| 463.60 | 21569.47(2) | 0.1702(2) | 10.47(3) | (iv) |

**Table 2**

| $\lambda_{head}$ (nm) | $\nu_0$ (cm$^{-1}$) | $B'$ (cm$^{-1}$) | $\Delta\nu_0^i$ (cm$^{-1}$) |
|---|---|---|---|
| 551.93 | 18117.30(3) | 0.1798(2) | 4.84(4) |
| 548.02 | 18246.78(3) | 0.1775(2) | -- |
| 538.55 | 18567.85(3) | 0.1705(3) | -- |
| 537.18 | 18614.89(2) | 0.1765(1) | 15.27(3) |
| 533.32 | 18749.44(2) | 0.1793(1) | 10.35(3) |
| 532.62 | 18773.94(2) | 0.1781(1) | -- |
| 514.62 | 19430.86(3) | 0.1737(2) | 13.35(4) |
| 475.66 | 21022.73(3) | 0.1661(2) | 22.03(4) |
| 466.92 | 21416.13(2) | 0.1658(2) | 16.06(3) |
| 459.54 | 21760.42(2) | 0.1658(1) | 9.38(3) |

**Table 3**

| Bands | (a) | (b) | (c) | (d) | (c) - (a) | (d) - (b) | (b) - (a) | (d) - (c) |
|---|---|---|---|---|---|---|---|---|
| (i) | -- | -- | 18150.1 | 18158.2 | -- | -- | -- | 8.1 |
| (i) | -- | -- | 18510.1 | 18515.6 | -- | -- | -- | 5.5 |
| (i) | 18763.3 | 18770.8 | 18832.1 | 18844.1 | 68.8 | 73.3 | 7.5 | 12.0 |
| (i) | 19103.9 | 19140.8 | 19174.3 | 19212.6 | 70.4 | 71.8 | 36.9 | 38.3 |
| (i) | 19451.5 | 19473.6 | 19524.1 | 19549.5 | 72.6 | 75.9 | 22.1 | 25.4 |
| (i) | -- | 19812.7 | 19854.3 | 19881.0 | -- | 68.3 | -- | 26.7 |
| (i) | 20147.1 | 20152.5 | 20216.2 | 20226.5 | 69.1 | 74.0 | 5.4 | 10.3 |
| (i) | -- | -- | 20584.7 | -- | -- | -- | -- | -- |
| (ii) | 19573.8 | -- | 19650.3 | -- | 76.5 | -- | -- | -- |
| (ii) | 19949.5 | 19966.7 | 20018.7 | 20040.2 | 69.2 | 73.5 | 17.2 | 21.5 |
| (ii) | -- | 20322.9 | 20378.0 | 20394.9 | -- | 72 | -- | 16.9 |
| (ii) | 20681.8 | 20693.7 | 20751.1 | 20764.7 | 69.3 | 71 | 11.9 | 13.6 |
| (ii) | 21038.2 | 21061.7 | 21108.1 | 21130.6 | 69.9 | 68.9 | 23.5 | 22.5 |
| (ii) | -- | -- | 21468.0 | -- | -- | -- | -- | -- |
| (iii) | -- | -- | 20276.9 | 20283.0 | -- | -- | -- | 6.1 |
| (iii) | 20568.0 | -- | 20634.8 | 20644.5 | 66.8 | -- | -- | 9.7 |
| (iii) | -- | -- | 20980.6 | -- | -- | -- | -- | -- |
| (iii) | -- | 21278.8 | 21333.9 | 21349.1 | -- | 70.3 | -- | 15.2 |
| (iii) | -- | 21621.0 | 21683.4 | 21690.4 | -- | 69.4 | -- | 7.0 |
| (iii) | -- | 21979.6 | 22039.5 | 22051.5 | -- | 71.9 | -- | 12.0 |
| (iv) | -- | -- | 20152.6 | -- | -- | -- | -- | -- |
| (iv) | -- | -- | 20510.9 | -- | -- | -- | -- | -- |
| (iv) | -- | 20824.3 | 20863.3 | 20897.0 | -- | 72.7 | -- | 33.7 |
| (iv) | -- | 21170.6 | 21205.3 | 21244.5 | -- | 73.9 | -- | 39.2 |
| (iv) | 21498.3 | 21528.3 | 21570.2 | 21599.3 | 71.9 | 71.0 | 30.0 | 29.1 |
| | 18043.4 | -- | 18118.2 | 18150.2 | 74.8 | -- | -- | 32.0 |
| | -- | 18198.2 | 18247.6 | 18274.3 | -- | 76.1 | -- | 26.7 |
| | -- | -- | 18568.5 | 18578.5 | -- | -- | -- | 10.0 |
| | -- | -- | 18615.5 | 18628.0 | -- | -- | -- | 12.5 |
| | 18671.8 | -- | 18750.4 | 18753.5 | 78.6 | -- | -- | 3.1 |
| | 18701.1 | 18705.3 | 18775.0 | 18779.9 | 73.9 | 74.6 | 4.2 | 4.9 |
| | 19361.4 | -- | 19431.6 | 19451.5 | 70.2 | -- | -- | 19.9 |
| | -- | -- | 21023.5 | 21038.2 | -- | -- | -- | 14.7 |
| | 21348.9 | 21358.7 | 21416.7 | 21426.4 | 67.8 | 67.7 | 9.8 | 9.7 |
| | 21690.3 | 21715.2 | 21761.1 | 21786.1 | 70.8 | 70.9 | 24.9 | 25.0 |

**Table 4**

|   | $^{58}$NiS | | $^{60}$NiS | |
|---|---|---|---|---|
| J | R(J) | P(J) | R(J) | P(J) |

**(a) 503.67 nm band**

| J | R(J) | P(J) | R(J) | P(J) |
|---|---|---|---|---|
| 1 |  | 19853.02 |  | 19838.76 |
| 2 |  | 19852.43 |  | 19838.23 |
| 3 |  | 19851.84 |  | 19837.75 |
| 4 |  | 19851.17 |  | 19837.01 |
| 5 |  | 19850.49 |  | 19836.22 |
| 6 | 19854.36 | 19849.84 | 19840.14 | 19835.59 |
| 7 | 19854.16 | 19848.98 | 19839.91 | 19834.77 |
| 8 | 19853.88 | 19848.01 | 19839.65 | 19833.78 |
| 9 | 19853.52 | 19846.97 | 19839.32 | 19832.78 |
| 10 | 19853.09 | 19845.87 | 19838.89 | 19831.66 |
| 11 | 19852.57 | 19844.67 | 19838.36 | 19830.48 |
| 12 | 19852.04 | 19843.44 | 19837.79 | 19829.19 |
| 13 | 19851.37 | 19842.08 | 19837.18 | 19827.86 |
| 14 | 19850.66 | 19840.75 | 19836.46 | 19826.50 |
| 15 | 19849.77 | 19839.24 | 19835.51 | 19824.99 |
| 16 | 19848.94 | 19837.56 | 19834.65 | 19823.39 |
| 17 | 19847.91 | 19835.71 | 19833.69 |  |
| 18 | 19846.88 | 19834.14 | 19832.64 |  |
| 19 | 19845.77 | 19832.32 | 19831.80 |  |
| 20 | 19844.60 | 19830.29 | 19830.65 |  |
| 21 | 19843.32 | 19828.76 | 19829.36 |  |
| 22 | 19841.94 | 19826.25 | 19828.06 |  |
| 23 | 19840.59 | 19824.65 | 19826.72 |  |
| 24 | 19839.14 |  | 19825.22 |  |
| 25 |  |  | 19823.68 |  |

**(b) 459.54 nm band**

| J | R(J) | P(J) | R(J) | P(J) |
|---|---|---|---|---|
| 1 |  | 21759.97 |  | 21742.25 |
| 2 |  | 21759.44 |  | 21741.69 |
| 3 |  | 21758.83 |  | 21741.10 |
| 4 |  | 21758.17 |  | 21740.47 |
| 5 | 21761.03 | 21757.40 | 21743.24 | 21739.63 |
| 6 | 21760.77 | 21756.51 | 21742.88 | 21738.74 |
| 7 | 21760.44 | 21755.47 | 21742.63 | 21737.67 |
| 8 | 21760.01 | 21754.43 | 21742.25 | 21736.63 |
| 9 | 21759.54 | 21753.25 | 21741.67 | 21735.44 |
| 10 | 21758.89 | 21752.07 | 21741.06 | 21734.16 |
| 11 | 21758.22 | 21750.70 | 21740.40 | 21732.74 |
| 12 | 21757.45 | 21749.24 | 21739.60 | 21731.30 |
| 13 | 21756.55 | 21747.72 | 21738.77 | 21729.89 |
| 14 | 21755.59 | 21746.21 | 21737.85 | |
| 15 | 21754.51 | 21744.44 | 21736.67 | |
| 16 | 21753.36 | 21742.62 | 21735.68 | |
| 17 | 21752.13 | 21740.69 | 21734.47 | |
| 18 | 21750.78 | 21738.71 | 21733.09 | |
| 19 | 21749.35 | 21736.41 | 21731.74 | |
| 20 | 21747.84 | 21734.27 | 21730.19 | |
| 21 | | 21732.15 | | |